\input{aipcheck}

\documentclass[
    ,final            
  ]
  {aipproc}
\usepackage{epsfig}
\layoutstyle{6x9}

\begin{document}

\title{Charge dynamics of $t$-$J$ model and anomalous bond-stretching
phonons in cuprates}

\author{P. Horsch}{
  address={Max-Planck-Institut f\"ur Festk\"orperforschung, D-70569 Stuttgart,
 Germany}
}

\author{G. Khaliullin}{
  address={Max-Planck-Institut f\"ur Festk\"orperforschung, D-70569 Stuttgart,
 Germany}
}

\begin{abstract}
The density response of a doped Mott-Hubbard insulator is discussed starting
from the $t$-$J$ model in a slave boson $1/N$ representation.
In leading order $O(1)$ the density fluctuation spectra  $N({\bf q},\omega)$
are determined by
an undamped  collective mode at large momentum transfer, in striking 
disagreement with results obtained by exact diagonalization, which
reveal a very broad dispersive peak, reminescent of strong spin-charge 
coupling. The $1/N$ corrections introduce the polaron character of the 
bosonic holes moving in a uniform RVB background.
The resulting $N({\bf q},\omega)$ captures all features observed in
diagonalization studies, fulfills the appropriate sum rules, and apart
from the broadening of the collective mode shows a new low energy feature 
at the  energy $\chi J+\delta t$ related to the polaron motion in the spinon
background. It is further shown that the low energy structure, which is
particularly pronounced in $(\pi,0)$ direction, describes the strong
renormalization and anomalous damping  of the highest bond-stretching phonons
in La$_{2-x}$Sr$_x$CuO$_4$.\\
\noindent
\centerline{
{\it Dedicated to Ferdinando Mancini on the occasion of his 60th birthday}}\\
\centerline{\it published in AIP Conference Proceedings 695: 
``Highlights in Condensed Matter Physics'', }\\
\centerline{ed. A. Avella et al. (AIP, New York, 2003) }

\end{abstract}

\maketitle


\section{Introduction}
High-temperature superconductors are doped Mott-Hubbard insulators,
therefore the density response is expected to be very different from
that of weakly correlated Fermi systems. 
The low-energy density response, i.e. in the frequency range inside
the Mott-Hubbard or charge transfer gap, is proportional
to the doping, i.e., described by transitions within the lower Hubbard band.
For a complete understanding of the physics ruling these systems both the spin 
and the charge response must be considered in the full momentum ${\bf q}$
and frequency ${\omega}$ domain.
The spin response has been and still is extensively explored, mainly 
stimulated by a wealth of experimental data provided by neutron scattering.
A similarly powerful experimental tool is missing for the charge response, with the
exception of optical conductivity, which however provides information only in the
limit ${\bf q} \rightarrow 0$.  
In the absence of solid experimental information about the detailed structure of
$N({\bf q},\omega)$, the best we can do to test a theory is the computer experiment.
This provides for small clusters
an unbiased answer how $N({\bf q},\omega)$ looks like, e.g., for the
$t$-$J$ model\cite{toh95,ede95}, and how this quantity depends on the
key parameters, namely the magnetic exchange interaction $J$ and the 
doping $\delta$.

In one dimension the Hubbard physics is basically simple
(although not mathematically), as it 
is characterized by charge and spin 
separation, which implies that the density response is basically that of
non-interacting spinless 
fermions, i.e., showing vanishing excitation energy at $q=0$ and $4 k_F$. 
This is even true for the 1D $t$-$J$ model away from $J=0$ and $2t$, where the
model is not exactly solvable\cite{toh95}. 
As we shall see the 2D model relevant for the cuprates 
shows a very different behavior compared to the 1D case.

Exact diagonalization studies for the $t$-$J$ model\cite{toh95,ede95}
have revealed that the dynamical density response $N({\bf q},\omega)$ 
for the 2D model  
is characterized by {\it strong spin-charge coupling}. 
These calculations show several features unexpected from the
point of view of weakly correlated fermion systems which have to be
explained by theory: (i) the strong
suppression of low energy $2k_F$ scattering in the density response,
(ii) a broad incoherent peak at high energy (several $t$) whose shape 
is rather insensitve to changes of hole
concentration and exchange interaction $J$, (iii) the very different
form of $N({\bf q},\omega)$ compared to the spin response function
$S({\bf q},\omega)$, which share common features in weakly correlated fermionic
systems, and finally (iv) there are at low energy ($\sim J$) some 
features that do depend on doping and $J$. It is worthwhile to note 
that finite temperature diagonalization studies\cite{jak00}
show only weak temperature dependence for $T < 0.3 t$ even at low energy.

While considerable analytical work has been done to explain the spin
response of the $t$-$J$ model only few authors analysed $N({\bf q},
\omega)$. Wang {\it et al.}\cite{wan91} studied collective excitations in the
density channel and found sharp peaks at large momenta
corresponding to free bosons. Similar results were obtained by Gehlhoff and
Zeyher\cite{geh95} using the X-operator formalism and by Foussats and
Greco\cite{greco02} using a path integral representation for X-operators. 
Lee {\it et al.}\cite{lee96} obtained a broad incoherent density fluctuation 
spectrum by considering a model of bosons coupled to a quasistatic
disordered gauge field, while Jackeli and Plakida\cite{jac99} employed the
memory function approach.

\begin{figure}[htb]
{\epsfig{file=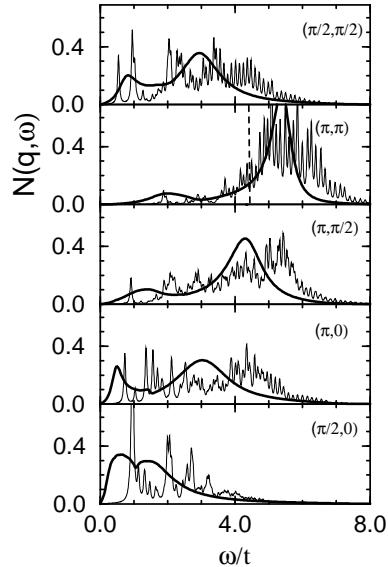,height=50mm,angle=-90}}
\caption{
Comparison of $N({\bf q},\omega)$ obtained by the slave-boson 
theory\cite{kha96} (solid lines) with diagonalization data\cite{toh95}
 for a periodic $4\times4$ cluster 
with $J/t=0.4$ and doping $\delta=0.25$. The dashed line in the $(\pi,\pi)$
spectrum indicates the $\delta$-function collective peak obtained when
polaron effects are neglected.  
}
\label{fig1}
\end{figure}

Starting from a slave-boson representation we
show that the essential features observed
in the numerical studies can be obtained in the framework of the
Fermi-liquid phase of the $t$-$J$ model at zero temperature\cite{kha96}. 
Our main
findings are: (i) at low momenta the main effect of strong
correlations is to transfer spectral weight from particle-hole
excitations into a pronounced collective mode. Because of the strong
damping of this mode (linear in $q$) due to the coupling to the spinon
particle-hole continuum, this collective excitation is qualitatively
different from a sound mode. (ii) At large momenta we find a strict
similarity of $N({\bf q},\omega)$ with the spectral function of a
single hole moving in a uniform RVB spinon background. 
In this regime $N({\bf q},\omega)$ consists of a broad peak at high
energy whose origin is the fast, incoherent motion of bare holes. 
(iii) The polaronic nature of dressed holes leads to the formation of a
second peak at lower energy, which is more pronounced in $(\pi,0)$
direction in agreement with diagonalization studies\cite{toh95,ede95}.

We show that
the anomalous renormalization of certain phonon modes as observed in inelastic 
neutron scattering provides a sensitive test of the pecularities of the 
low-energy density response.  In this contribution we shall analyse the strong,
doping-dependent renormalization of the highest breathing phonon modes
which is a generic feature in the high-$T_c$ compounds.


\section{Slave boson theory of density response}

We start from the $N$-component generalization of the slave-boson $t$-$J$
Hamiltonian\cite{kot88,wan91} $H_{tJ}=H_t + H_J$, which is obtained by replacing
the constrained electron creation operators 
${\tilde c}^+_{i,\sigma}=c^+_{i,\sigma}(1-n_{i,-\sigma})
\rightarrow f^+_{i,\sigma}h_i$ :
\begin{equation}
H_t = -\frac{2t}{N}\sum_{<i,j>\sigma} (f^+_{i\sigma} h^+_j
h_i f_{j\sigma} + h.c.),
\end{equation}
\begin{equation}
H_J = \frac{J}{N}\sum_{<i,j>\sigma\sigma'} f^+_{i\sigma}
f_{i\sigma'}  f^+_{j\sigma'} f_{j\sigma}
(1-h^+_i h_i)(1-h^+_j h_j),
\end{equation}
where $f^+_{i,\sigma}$ is a fermionic (spinon) operator,
 $\sigma =1,\cdots,N$ is the fermionic flavor index, and 
$h_i$ denotes the bosonic holes. These operators obey standard 
commutation rules, yet
the number of these auxiliary particles must obey the constraint  
$\sum_{\sigma} f^+_{i\sigma}f_{i\sigma} +h^+_ih_i=N/2$.
The original $t$-$J$ model is recovered for $N=2$.

The slave boson parametrization provides a straightforward description
of the strong suppression of density fluctuations of constrained
electrons through the representation of the density response in terms
of a dilute gas of bosons. A common treatment of model (1) is the
density-phase representation (``radial'' gauge\cite{rea83}) of the
bosonic operator $h_i=r_i \exp(i\theta_i)$ with the subsequent
$1/N$-expansion around the Fermi-liquid saddle point. While this gauge
is particularly useful to study the low energy and momentum
properties, it is not very convenient for the study of the density
response in the full $\omega$ and ${\bf q}$ space. Formally the latter
follows in the radial gauge from the fluctuations of $r_i^2$. However, if one
considers for example convolution type bubble diagrams, one realizes
that their contribution to the static structure factor is correctly of
order $1/N$, but is not proportional to the density 
of holes $\delta$ as it should be.
According to Arrigoni et al\cite{arr94} such unphysical
results originate from a large negative pole contribution in the 
$\langle r_{-\bf q}r_{\bf q}\rangle_{\omega}$ 
Green's function of the real field $r$, which is
hard to control by a perturbative treatment of phase fluctuations. 

We follow therefore Popov\cite{pop83} using the density-phase
treatment only for small momenta $q<q_0$, while keeping the original
particle-hole representation of  the density operator, $h^+h$, at
large momenta. More precisely $h_i=r_i\exp(i\theta_i) + b_i$, where
$b_i=\sum_{|{\bf q}|>q_0} h_{\bf q}\exp(i{\bf q}{\bf R}_i)$. The
cutoff $q_0$ is introduced dividing ``slow'' (collective) variables
represented by $r$ and $\theta$ from ``fast'' (single-particle)
degrees of freedom $b_i$ and $b_i^+$. 
As discussed by Popov\cite{pop83} this ``mixed''
gauge is particularly useful for finite temperature studies to
control infrared divergences. We start formally with the ``mixed'' gauge
and keep only terms of order $\delta$ and $1/N$ in the bosonic self
energies. In this approximation our zero temperature calculations
become quite straightforward: The cutoff $q_0<\delta$ actually does
not enter in the results and we arrive finally at the Bogoliubov
theory for a dilute gas of bosons moving in a fluctuating spinon
background. 

The Lagrangian corresponding to the model (1) is then given by
(the summation over $\sigma$ is implied)
\begin{eqnarray}
L&=&\sum_{i} \Bigl( {f^+_{i\sigma}}
  \bigl({{\partial}\over{\partial\tau}}-\mu_f\bigr) f_{i\sigma}
  +b^+_i \bigl( {{\partial}\over {\partial\tau}}-\mu_b\bigr) b_i\Bigr) 
  +H_{t}+H_J \nonumber \\
& &+\frac{i}{\sqrt{N}}\sum_{i}\lambda_i\Bigl(f^+_{i\sigma}
   f_{i\sigma}+(r_i+b_i^+)(r_i+b_i)-\frac{N}{2}\Bigr),
\end{eqnarray}
\begin{equation}
H_t=-\frac{2t}{N}\sum_{<ij>}f^+_{i\sigma}f_{j\sigma}
   \bigl( b^+_jb_i+r_ir_j+r_jb_i+b_j^+r_i\bigr)+h.c.
\end{equation}
Here the $\lambda$ field is introduced to enforce the constraint, and
$\mu_f,\mu_b$ are fixed by the particle number equations
$\langle n_f\rangle =\frac{N}{2} (1-\delta)$ and
$\langle r^2_i+b^+_ib_i\rangle =\frac{N}{2}\delta$, respectively.
The uniform mean field solution $r_i=r_0\sqrt{N/2}$ leads in the large
$N$ limit to the renormalized narrow fermionic spectrum 
$\xi_{\bf k}=-z\tilde t \gamma_{\bf k}-\mu_f$, with $\tilde t =J\chi+t\delta$, 
$\gamma_{\bf k}=\frac{1}{2}(\cos{k_x}+\cos{k_y})$,
$\chi=\sum_{\sigma}\langle f^+_{i\sigma} f_{j\sigma}\rangle /N$, and $z=4$ the
number of nearest neighbors.
In the $N=\infty $ limit $\chi_{\infty}\simeq 2/\pi^2$ is given by that of
free fermions, while for the original $t$-$J$ model its value should
be larger\cite{zha88} due to Gutzwiller projection.
In the following $\chi=\frac{3}{2}\chi_{\infty}$ will be used. 
Distinct from the finite-temperature gauge-field theory of
Nagaosa and Lee\cite{nag90} the bond-order phase fluctuations acquire
a characteristic energy scale in this approach\cite{wan91}, and the
fermionic (``spinon'') excitations can be identified with Fermi-liquid
quasiparticles.
The mean field spectrum of bosons is
$\omega_{\bf q}=2z\chi t(1-\gamma_{\bf q})$. Thus the effective mass of holes
$m^0_h\propto 1/t$ is much smaller than that of the spinons. 

Due to the diluteness of the bosonic subsystem, $\delta\ll1$, the
density correlation function 
$\chi_{{\bf q},\omega}=\langle \delta n^h \delta n^h\rangle_{{\bf q}\omega}$
is mainly given by the condensate induced
part which is represented by the Green's function 
$\langle (b^+_{\bf q}
+b_{\bf -q})(b_{\bf q}+b^+_{\bf -q})\rangle_{\omega}$ 
for $q>q_0$, 
and $2\langle r_{\bf -q}r_{\bf q}\rangle_{\omega}$ for  $q<q_0$,
respectively: 
\begin{equation}
\chi_{{\bf q}\omega}\simeq\frac{N}{2}r_0^2\Bigl(
\langle (b^+_{\bf q}
+b_{\bf -q})(b_{\bf q}+b^+_{\bf -q})\rangle_{q>q_0}
+2\langle r_{\bf -q}r_{\bf q}\rangle_{q<q_0} \Bigr).
\end{equation} 
The $1/N$ self-energy corrections to these functions are calculated in
a conventional way\cite{rea83,kot88} expanding
$r_i=(r_0\sqrt{N}+(\delta r)_i)/\sqrt{2}$ and considering Gaussian
fluctuations  around the mean field solution. Neglecting all terms of
order $\delta/N$ and $q^2_0/N$, only one relevant $1/N$ contribution
remains which corresponds to the dressing of the slave-boson Green's
function by spinon particle-hole excitations. Within this
approximation and at zero temperature no divergences occur at low
momenta, thus one can take the limit $q_0\rightarrow0$. The final
result for the dynamic stucture factor ({\it normalized by the hole
density}) is: 
\begin{equation}
N_{{\bf q},\omega}=\frac{2}{\pi} Im \Bigl( \bigl( \omega_{\bf q}a 
 +S^{(1/N)}_{{\bf q},\omega}-\mu_b\bigr)/D_{{\bf q},\omega} \Bigr),
\end{equation}
\begin{equation}
D_{{\bf q},\omega}= \bigl( \omega_{\bf q}a 
 +S^{(1/N)}_{{\bf q},\omega}-\mu_b\bigr) \bigl(\omega_{\bf q} 
 +S^{(1)}_{{\bf q},\omega}+S^{(1/N)}_{{\bf q},\omega}-\mu_b\bigr)
 -\bigl(\omega a -A^{(1/N)}_{{\bf q},\omega}\bigr)^2.
\end{equation}
The origin of the contribution
\begin{equation}
S^{(1)}_{{\bf q},\omega}=ztr_0^2
\bigl(\frac{(1+\Pi_2)^2}{\Pi_1}-\Pi_3\bigr)_{{\bf q},\omega},
\label{S1}
\end{equation}
\begin{equation}
\Pi_m=zt\sum_{\bf k}\frac{n(\xi_{\bf k})-n(\xi_{\bf k+q})}
 {\xi_{\bf k+q}-\xi_{\bf k}-\omega-i0^+}(\gamma_{\bf k}
 +\gamma_{\bf k+q})^{m-1},\nonumber
\end{equation}
is the indirect interaction of bosons via the spinon band due to the
hopping term (which gives $\Pi_3$ in (4)) and due to the coupling to
spinons via the constraint field $\lambda$.
The latter channel provides a repulsion between bosons, making 
$S^{(1)}(\omega=0)$ positive and therefore ensuring the stability of
the uniform mean-field solution. 

The $1/N$ self energies $S^{(1/N)}$
and $A^{(1/N)}$ are essentially a single boson property. They are
given by the symmetric and antisymmetric combinations 
(with respect to $\omega+i0^+\rightarrow -\omega-i0^+$)
of the self energy
\begin{equation}
\Sigma^{(1/N)}_{{\bf q},\omega}=\frac{4}{N}
\sum_{|{\bf k}|<k_F<|{\bf k'}|} 
(zt\gamma_{\bf k'-q})^2 G^0_{\bf q+k-k'}  
(\omega+\xi_{\bf k}-\xi_{\bf k'}).
\end{equation}
Here $G^0_{\bf q}(\omega)=(\omega-\omega_{\bf q}-
\Sigma^{(1/N)}_{{\bf q},\omega}+\mu_b)^{-1}$ 
is the Green's function for a single  boson moving in a 
uniform RVB background. 
Although in the context of $1/N$ theory the $G^0$ function in (5)
should be considered as a free propagator, we shall use here the
selfconsistent polaron picture for a single hole\cite{kan89}.
This is crucial when comparing the theory for $N=2$ with
diagonalization studies. Finally, the constants $a$ and $\mu_b$ in (3)
are given by $(1-t r^2_0/{\tilde t})$ and 
$S^{(1/N)}(\omega={\bf q}=0)$, respectively.
The parameter $r_0^2$ in Eq.\ref{S1}, which formally corresponds to the
condensate fraction in our theory, is determined selfconsistently from 
$r_0^2=\delta-\sum_{{\bf q}\neq 0}\tilde n_{\bf q}$. 
The momentum
distribution $\tilde n_{\bf q}=\langle b^+_{\bf q}b_{\bf q}\rangle$ is
calculated from the corresponding bosonic Green's function for finite
hole-density. The interactions implied in the polaron formation lead to
a significant reduction of the number of bosons in the
condensate\cite{kha96}, and affects the balance of the  selfenergies
$S^{(1)}$ and $S^{(1/N)}$ in Eq. (7).
\vspace*{3mm}
\begin{figure}[htb]
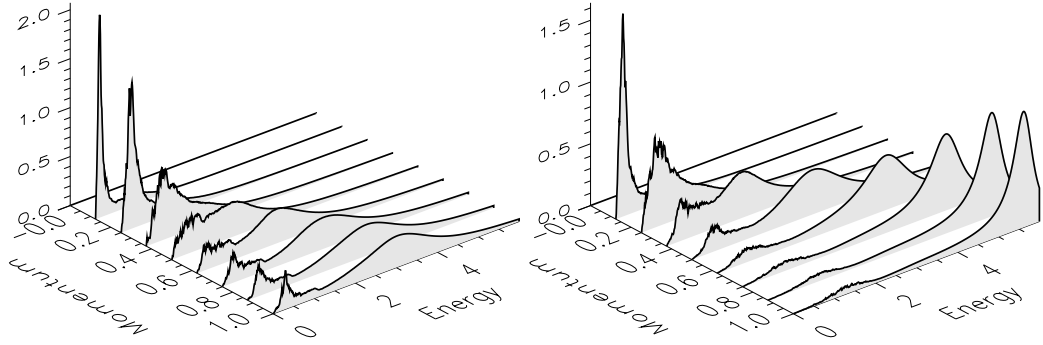

\epsfig{file=fig2a.epsi,height=45mm,angle=0} 
\epsfig{file=fig2b.epsi,height=45mm,angle=0}
\vspace*{3mm}
\caption{Density fluctuation spectra $N({\bf q},\omega)$ for
$J/t=0.4$ and $\delta=0.15$ along $(\pi,0)$ (left) and
$(\pi,\pi)$ directions (right). Energy in units of $t$.
}
\label{fig2}
\end{figure}

The structure of  $N({\bf q},\omega)$ (6)
in the small $\omega,{\bf q}$ limit  is mainly
controlled by the interaction of bosons represented by the $S^{(1)}$
term ($\propto r_0^2$), while the internal polaron structure of the boson 
determined by $S^{(1/N)}$ is less important.
$N({\bf q},\omega)$ consists of a weak spinon particle-hole continuum
with cutoff $\propto v_F q$, and a very pronounced linear collective mode
which nearly exhausts the sum rule. The velocity of this
mode is always somewhat smaller than the spinon Fermi velocity, $v_s\leq v_F
\simeq z\tilde t$, which implies a strong 
damping $\propto\omega$ (or $q$) of this mode (Fig.\ref{fig2}).

\begin{figure}[ht]
{\epsfig{file=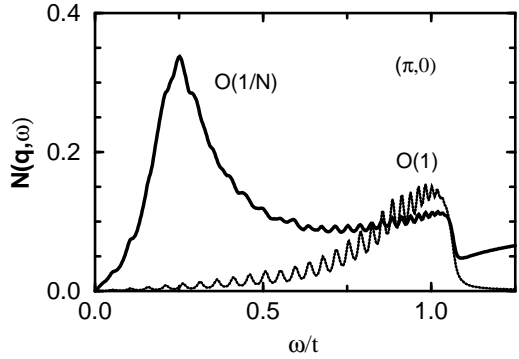,height=70mm,angle=-90}}
\caption{
Low energy density response at $(\pi,0)$ for $J=0.3$ and 
$\delta=0.18$. The peak centered at $\sim(\chi J+\delta t)$
due to the polaron motion appears in the $1/N$ order
on top of the spinon particle-hole continuum with edge at energy
$z(\chi J+\delta t)$. 
Small oscillations in the data result from the discretization of the
energy scale in the numerical solution and have no physical meaning.
}
\label{fig3}
\end{figure}

\begin{figure}[ht]
{\epsfig{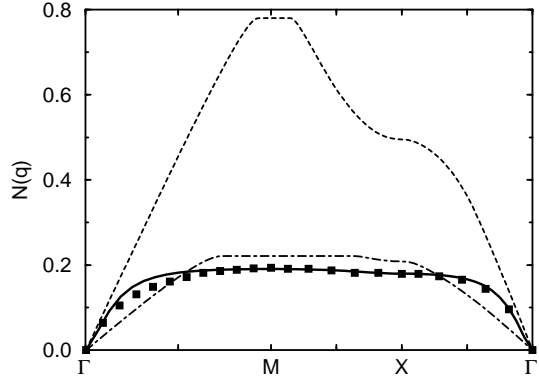}}
\caption{Comparison of static structure factor $N({\bf q})$ (solid line)
calculated for $J=0.4$ with the result for a Gutzwiller projected
wave function\cite{gros94} (squares) and spinless fermions (dash-dotted)
for $\delta=0.213$.  The comparison with the result obtained for 
non-interacting fermions with spin (dashed line) shows the large reduction
of the density response due to the constraint.
}
\label{fig4}
\end{figure}

The density response  $N({\bf q},\omega)$ (Fig.2) at large momenta,
$q>\delta$, which we can compare with diagonalization results
(Fig.1), is
dominated by the properties of the single boson selfenergy $S^{(1/N)}$.
The calculated density response of the $t$-$J$ model has
three characteristic features on different energy scales:
(i) 
The main spectral weight of the excitations at large momenta is
located in an energy region of order of several $t$. This high energy
peak is very broad and incoherent as a result of the strong coupling
of bosons to low-energy spin excitations. 
The position of
this peak and its shape are rather insensitive to the ratio $J/t\leq1$ 
in agreement with conclusions of \cite{toh95,ede95}.
This is simply due to the fact that the high-energy properties of the
$t$-$J$ model are controlled by $t$.
(ii)
The theory predicts also a second peak at lower energy 
which is more pronounced in the direction $(\pi,0)$ (see also Fig.3), 
while its weight is strongly suppressed for ${\bf q}$ near $(\pi,\pi)$. 
The origin of this excitation is due to the formation of a
polaron-like band of dressed bosons. The relative weight of this
contribution  increases with $J$ as a result of the increasing
spinon bandwidth. 
(iii)
In addition there is the spinon particle-hole continuum which is
generated by  $S^{(1)}$ with relatively small weight ($\propto \delta^2$). 
At $(\pi,0)$ the high energy cutoff of the spinon continuum
is at $z(\chi J+\delta t)$ (Fig.3),
while the polaron peak is at about  $(\chi J+\delta t)$.
We note that the polaron peak in  $N({\bf q},\omega)$ at optimal doping
is in the same energy range as the high energy longitudinal
optical phonons in cuprates, and as we shall discuss below leads
to a strong damping and anomalous renormalization  of these modes for certain
wave numbers.

Figure 4 provides a comparison of the static charge structure factor 
$N({\bf q})$ with that of a Gutzwiller projected wave function and with 
that for uncorrelated electrons. 
The latter comparison shows the strong suppression of 
the density response due to correlations.  
Finally we note that our calculated structure $N({\bf q})$ factor satisfies
the sum rule for constrained fermions  $\sum_qN({\bf q})=\delta(1-\delta)$
within $1\%$ \cite{kha96}.


\section{Renormalization of breathing phonons}

\begin{figure}[ht]
{\epsfig{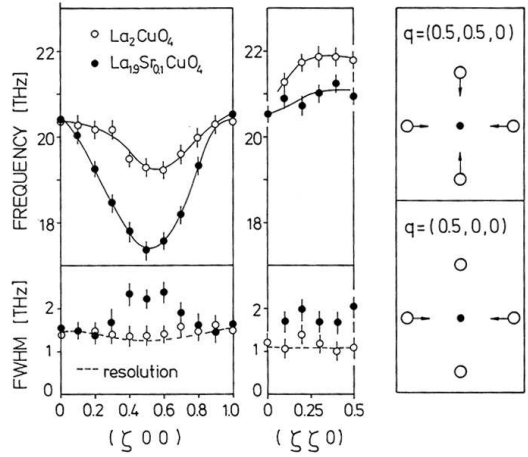}}
\caption{
Doping dependence of the high energy bond-stretching phonons
along $(\zeta,0,0)$ and $(\zeta,\zeta,0)$ directions
for undoped La$_2$CuO$_4$ (open circles) and  
La$_{1.85}$Sr$_{0.15}$CuO$_4$ (filled circles) as obtained by neutron
scattering. The lower panel gives the line width at half maximum. 
The displacements of oxygen ions (a) for the ${\bf q}=(\pi,\pi)$ breathing
phonon distortion and (b) for  the $(\pi,0)$ half-breathing mode are
indicated by arrows in the right panel. 
Here $(\pi,0)$ corresponds to (0.5,0,0) reduced lattice units.
(reprinted from Pintschovius und Reichardt\cite{pin94} with permission
from World Scientific)
}
\label{fig5}
\end{figure}

Inelastic neutron scattering experiments on high-$T_c$ superconductors
have shown that in particular the highest energy longitudinal
optic phonon branch near $(\pi,0)$ softens and broadens strongly as holes
are doped into the insulating parent compound. Whereas the corresponding
breathing mode at $(\pi,\pi)$ shows much smaller softening and no
anomalous broadening (Fig.\ref{fig5}). 
This effect appears to be generic for cuprates
and detailed neutron scattering studies have been reported for 
La$_{2-x}$Sr$_x$CuO$_4$ \cite{pin91,pin94,mcq96,pin99}, 
YBa$_2$Cu$_3$O$_{6+x}$\cite{pin94,rei96,pin02} and also for 
Bi-based cuprates\cite{ren89}.
This certainly shows that there are phonons which couple rather strongly to
the charge carriers that contribute to superconductivity, 
and one may ask
what is special about the interaction of strongly correlated electrons and phonons.

\vspace*{5mm}

\begin{figure}[ht]
{\epsfig{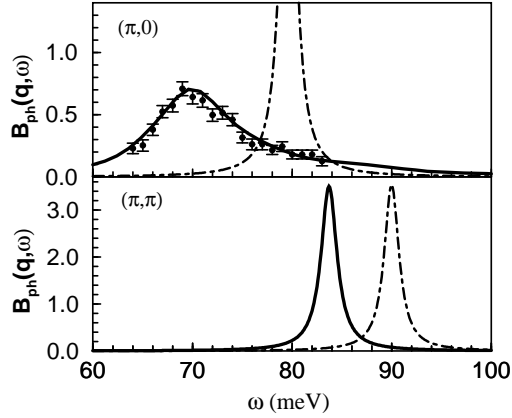}}
\caption{Calculated phonon spectral function $B_{ph}$ for $(\pi,0)$
and $(\pi,\pi)$ breathing phonons for $\delta=0.15$, $t=0.4$ eV, $J=0.12$ eV
and $\xi=0.25$ (solid lines)
compared to undoped system (dash-dotted lines) and the
inelastic neutron scattering data\cite{mcq96} for La$_{1.85}$Sr$_{0.15}$CuO$_4$.
(reprinted from Khaliullin and Horsch\cite{kha97} 
with permission from Elsevier)
}
\label{fig6}
\end{figure}
The renormalization of these 
phonons  can be calculated in the framework of the
$t$-$J$ model, since the breathing oxygen motion of 
these modes (see Fig.\ref{fig5}) modulates the energy of a hole in a
Zhang-Rice singlet state, and therefore couples directly to the density
of doped holes.
Expanding the Zhang-Rice energy  $E_{ZR}=8\frac{t_{pd}^2}{\Delta\epsilon}$
with respect to the oxygen displacements 
$u_{\alpha}^i=u_{\alpha}(R_i+\delta^O_{\alpha})$, 
$\alpha=x,y$, of the four O-neighbors at $R_i+\delta^O_{\alpha}$
around the Cu-hole at $R_i$ yields
the linear electron-phonon coupling \cite{kha97}
\begin{equation} 
H_{e-ph}=
g\sum_i(u^i_{x}-u^i_{-x}+u^i_{y}-u^i_{-y}) h^+_ih_i.
\end{equation}
We assume that the resonance integral obeys the Harrison relation 
$t_{pd}\propto  r_0^{-7/2}$, where $r_0$ is the Cu-O distance, and obtain
$g=7 E_{ZR}/ 4 r_0$, i.e., $g\approx 2$eV/\AA. 
The lattice part of the Hamiltonian is determined
by the force constant $K\approx 25$eV/\AA$^2$ for the longitudinal O-motion.
Due to the structure of $ H_{e-ph}$ the breathing modes couple directly
to $\chi_{{\bf q},\omega}$. 

We have studied the renormalization of the
phonon Green's functions along $(\pi,0)$ and
$(\pi,\pi)$ directions 
\begin{equation}
D^{ph}_{{\bf q},\omega}=\frac{\omega_{{\bf q},0}}
{\omega^2-\omega^2_{{\bf q},0}(1-\alpha_{\bf q}
\chi_{{\bf q},\omega})}, 
\end{equation}
where $\omega_{{\bf q},0}$ is the bare phonon frequency, i.e. measured in the undoped
parent compound, and $\alpha_{\bf q}=\frac{4g^2}{K}(\sin^2{q_x/2}
+\sin^2{q_y/2})$. Based on the parameters of the $pd$-model we estimate
for the dimensionless coupling constant $\xi=g^2/tK \sim 0.3 - 0.5$.
The coupling constant $\alpha_{\bf q}^2$ vanishes at the $\Gamma$ point and becomes 
maximal at the zone edges. This explains the marginal changes of renormalized
phonon frequency $\omega_{\bf q}$ at ${\bf q}=(0,0)$. The particularly strong 
increase of the renormalization for ${\bf q}$ along $(\pi,0)$ is a combined 
effect of the strong increase of   $\alpha_{\bf q}^2$ and the low energy polaron
structure in the density response.

\begin{figure}[ht]
{\epsfig{file=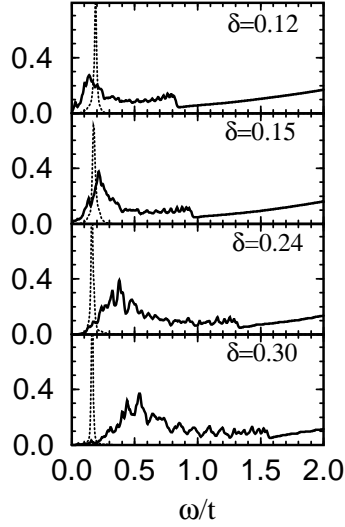,height=45mm,angle=-90}}
\caption{
Doping dependence of low-energy density response at $(\pi,0)$ (solid lines).
As a consequence of the scaling of the polaron structure
$\propto(\chi J + \delta t)$ there is a strong change 
in the renormalization and damping of the $(\pi,0)$ optical
phonon (dashed lines), which is at $\omega_0=0.2 t$ in the undoped system.  
}
\label{fig7}
\end{figure}

Figure \ref{fig6}, calculated for typical parameters for cuprate superconductors,
shows the strong renormalization  of the  
$(\pi,0)$ half-breathing mode for La$_{1.85}$Sr$_{0.15}$CuO$_4$
with a twice as large shift as for the $(\pi,\pi)$ breathing phonon.
The large damping of the $(\pi,0)$ phonon results from the 
hybridization with the large polaron
peak in $N({\bf q},\omega)$ at this momentum
and is consistent with the experimental data\cite{mcq96}. The phonon energies
of the undoped parent compound $\omega_{{\bf q},0}=80 (90)$ meV for $(\pi,0)$ and
$(\pi,\pi)$, respectively, are taken from Ref.\cite{pin94}.

The strong doping dependence of this effect, shown in Fig.\ref{fig7}, is due to
the scaling $\propto(\chi J + \delta t)$ of the polaron peak position 
in $N({\bf q},\omega)$.  
In particular our results imply that the large damping of the $(\pi,0)$
optical phonon should disappear at larger doping concentrations.

\section{Summary}
We have outlined a 1/N slave-boson theory for the density response
of the $t$-$J$ model, which explains the data obtained by exact
diagonalization. We demonstrated that the predicted
low energy polaron structure in the density response, which is particularly
pronounced along $(\pi,0)$, explains the anomalous doping induced
line width and shift of the longitudinal planar $(\pi,0)$ phonon.
The energy of the polaron peak is determined by the spinon energy
scale, therefore we predict a nontrivial doping dependence for
the phonon renormalization. In that respect 
further neutron scattering studies of the doping
dependence of phonons would provide a sensitive test for the low energy density
response as well as for the spin structure in the different doping regimes.


\begin{theacknowledgments}
We are grateful to L. Pintschovius for several illuminating discussions about
the interpretation of neutron scattering data and for permission to reprint
Figure 5.
\end{theacknowledgments}


\bibliographystyle{aipprocl} 



\end{document}